\documentstyle[12pt] {article}
\begin {document}
\parindent=15pt
\begin{center}
\vskip 1.5 truecm

{\bf QUARK MODEL AND NEUTRAL STRANGE SECONDARY PRODUCTION BY
NEUTRINO AND ANTINEUTRINO BEAMS} \\
\vspace{.5cm}
I.N.Erofeeva, V.S.Murzin \\
\vspace{.5cm}
Institute of Nuclear Physics, Moscow State University, \\
Moscow, Russia \\

\vspace{.5cm}
V.A.Nikonov, Yu.M.Shabelski \\
\vspace{.5cm}
Petersburg Nuclear Physics Institute, \\
Gatchina, St.Petersburg 188350 Russia \\
\end{center}
\vspace{1cm}

\begin{abstract}
The experimental data on $K^0$ and $\Lambda$ production by $\nu$ and
$\bar{\nu}$ beams are compared with the predictions of quark model
assuming that the direct production of secondaries dominates.
Disagreement of these predictions with the data allows one to suppose
that there exists considerable resonance decay contribution to the
multiplicities of produced secondaries.
\end{abstract}
\vspace{3cm}

%E-mail: $\;$ nikonov@thd.pnpi.spb.ru \\

%E-mail: $\;$ shabelsk@thd.pnpi.spb.ru \\
\newpage

\section{Introduction}
It is well-known that in soft hadron-hadron collisions the production
of resonances gives an important contribution to the multiplicity
of "stable" secondaries (such as pions, kaons, etc.). For example
in the additive quark model the probability of "direct" production
of secondary hadron having spin $J$ is proportional to the factor
$2J + 1$ that means the main parts of pions, kaons, etc. are
produced via decay of vector, tensor and higher spin resonances.
These results are in reasonable agreement \cite{AKNS} with the data in
soft hadron-hadron collisions.

However, the information about the role of resonance production
in hard processes is not sufficient. The mechanisms of multiparticle
production in soft and hard processes can be different. So in the
present paper we will consider the role of resonances in the neutral
strange secondary production in the deep inelastic interactions of high
energy neutrino and antineutrino with protons and neutrons.

\section{Experimental data}
For comparison with the Quark Model predictions (QM) the
experimental data of E632 Collaboration had been used \cite{E632}.
The experiment was done at the Fermilab Tevatron. The detector
was the 15-ft. bubble chamber filled with a liquid neon-hydrogen
mixture which also served as the target. The bubble chamber was
exposed to a neutrino beam. The neutrino beam was produced by
the quadrupole triplet train, which focused secondary particles
produced by the interactions of 800 GeV protons from the
Tevatron. 

The data sample consists of 6459 events ( 5416 - $\nu - Ne$ 
interactions and 1043 - $\bar{\nu} - Ne$ interactions).
The neutrino interactions with a single nucleon were picked out
by using the criterion of selection of the interactions with the
peripheral nucleon or the neutrino interactions without the
intranuclear cascades such as the mass of the target \cite{CSC}.
The neutrino-nucleon interactions could be selected into 
neutrino-proton and neutrino-neutron interactions by using
the total charge of the hadronic system (Table 1).
This material was used for the determination of the numbers of generated
$K^0$ and $\Lambda$ particles as well their parts in the different
groups of the events (Table 2).
In the data sample of vee of the Table 2 it is not taken into consideration
the corrections for losses of $K^0$ and $\Lambda$ particles caused
by the methodical sources (the limited volume of the bubble
chamber, scanning and fitting efficiency, etc.) \cite {E632}.
Nevertheless the weighted coefficients, taking into account these
effects, must not be distinguished for the neutrino-proton and 
the  neutrino-neutron interactions.

\section{Quark model predictions}
We will consider only events with charged current interactions.
In the case of interactions with sea quarks every type of particle and
antiparticle are produced practically in the same proportion
independently on their isospin projection (say, we expect the equal
multiplicities of $K^+$, $K^0$, $\bar{K}^0$ and $K^-$). However
the secondaries produced with comparatively large 
negative
Feynman-$x$
($x_F$) in the laboratory frame
, in the target fragmentation region, 
should contain valence quarks of the
target nucleon, so different kinds of kaons should be produced with
different probabilities. For the model prediction we will use the fact
that neutrino interacts with valence $d$-quark which transfers into
$u$-quark whereas antineutrino interacts with $u$-quark which transfers
into $d$-quark. So we have the following configurations:
\begin{equation}
\nu p \rightarrow uu + u' \;,
\end{equation}
\begin{equation}
\bar{\nu} n \rightarrow dd + d' \;,
\end{equation}
\begin{equation}
\bar{\nu} p \rightarrow ud + d' \;,
\end{equation}
\begin{equation}
\nu n \rightarrow ud + u' \;.
\end{equation}
Here $q'$ means the fast quark in the laboratory frame which absorbs
$W$-boson and determines the fragmentation in the current region and
another two quarks determine the fragmentation of valence remnant into
secondaries with comparatively large $x_F$ in the target hemisphere.

One can see from Eqs. (1)-(4) that, say, direct production of $K^0$
($d\bar{s}$) with comparatively large $x_F$ should be suppressed in
the process (1), where there are no valence $d$-quarks, in comparison
with another reactions. In the process (2), where there are two valence
$d$-quarks, it should be about two times larger than in the cases of
Eqs. (3) and (4).  However, if a significant part of $K^0$ can be
produced via decay of $K^+(890)$ and $K^0(890)$, the yields of $K^0$
with large $x_F$ can be more or less equal in all considered processes.

A similar situation appears in the case of secondary $\Lambda$-baryon
production with large $x_F$. The direct $\Lambda$ (containing two
initial valence quarks, $u$ and $d$) can be produced with equal
probabilities in the processes (3) and (4) and their production should
be suppressed in reactions (1) and (2). However in the case of
$\Lambda$ production via $\Lambda \pi$ decay of isotriplet resonance
$\Sigma (1385)$ the multiplicities of large-$x_F$ $\Lambda$ should be
of the same order in all reactions (1)-(4).

\section{Results}
Here we compare the experimental results on neutral kaons and
$\Lambda$-hyperon production by $\nu$ and $\bar \nu$ beams with quark
model predictions. The quark model predictions for the multiplicities
of strange secondaries assuming only direct production of a kaon
containing one valence quark of incident target nucleon and direct
production of $\Lambda$ containing two valence quarks of target nucleon
are presented in Table 2. Here $w_K$ and $w_{\Lambda}$ are the
probabilities of $K^0$ and $\Lambda$ production in the processes of
fragmentation (or recombination) of one and two valence quarks of the
target nucleon, respectively. Let us repeat again that in the case of
large contributions of resonance decay the multiplicities of $K^0$
and $\Lambda$ can be more or less equal (the exact values of their
ratios are model dependent).

One can compare the presented predictions with the experimental
multiplicities of $K^0$ with 
$x_F < -0.2$ and $\Lambda$ 
with $x_F < -0.4$.
 
It is clear that the data for 
the both 
$K^0$ 
and $\Lambda$ 
production do
not agree with the presented predictions for direct mechanism of
secondary production. 
Say, the multiplicity of $K^0$ in $\bar{\nu} n$ interactions should be 
equal to the sum of their multiplicities in $\nu n$ and $\bar{\nu} p$
interactions, i.e. 
$\simeq 0.005 \pm 0.002$ 
that is in disagreement with
the experimental value.
The most natural explanation is a large
resonance contribution to the multiplicities of neutral strange
secondaries which changes the predictions depending on the model of
resonance contributions.

\section{Conclusion}
We compare the experimental data on $K^0$ and $\Lambda$ production
by $\nu$ and $\bar{\nu}$ beams on the proton and neutron targets with
the predictions of quark model assuming that their direct production
dominates. Disagreement of these predictions with the data allows us to
suppose that there exists considerable resonance decay contribution to
the multiplicities of produced secondaries. Unfortunately the
experimental statistics are not large enough for numerical estimations.

This work is supported in part by grants RFBR 96-15.96764, NATO
OUTR. LG 971390 and RFFI 99-02-16578.
%%
%\end{document}

\newpage
\begin{table}[h]
\caption{\label{table:t1} The experimental data from E 632.}
\centering
\begin{tabular}{||c||c|c|c|c||} 
\hline
\hline
Reaction & $\nu(\bar{\nu})-Ne$ & $\nu(\bar{\nu})-N$ & $N(K^0)$  
& $N(\Lambda)$ \\
\hline
\hline
$\nu p$        & 5416   & 739  & 47  & 15 \\ 
\cline{1-1}
\cline{3-5}
$\nu n$        &        & 1273 & 84  & 38 \\ 
\hline
$\bar{\nu} p$  & 1043   & 282  & 20  & 7 \\
\cline{1-1}

\cline{3-5}
$\bar{\nu} n$  &        & 179  & 7   & 8 \\ 
\hline
\hline
\end{tabular}
\end{table}

\begin{table}[h]
\caption{\label{table:t2}The comparison of quark model (QM)
predictions for the multiplicities
of directly produced $K^0$ and $\Lambda$ at large $x_F$ with the
experimental data.} 

\centering
\begin{tabular}{||c||c|c||c|c||} 
\hline
\hline
Reaction & $K^0$ (QM) & $K^0$ (exp)
& $\Lambda$ (QM) & $\Lambda$ (exp) \\ 
    & & $x_F<-0.2$ & & $x_F<-0.4$ \\
\hline 
\hline
$\nu p$    & -   & $0.008 \pm 0.003$  & -  
& $0.004 \pm 0.002$  \\ 
$\nu n$    & $w_K$   & $0.005 \pm 0.002$ & $w_{\Lambda}$  
& $0.011 \pm 0.003$ \\
$\bar{\nu} p$  &  $w_K$  & $0.004 \pm 0.004$ & $w_{\Lambda}$ 
& $0.004 \pm 0.004$ \\
$ \bar{\nu} n$  & $2w_K$   & $0$ & - 
& $0$ \\   
\hline
\hline

\end{tabular}
\end{table}

\end{document}